\documentclass[a4paper,11pt]{article}
\usepackage{pos}
\usepackage{float}

\title{Gamma-Ray and AntiMatter Survey(GRAMS) experiment}

\author*[a]{J.C Zeng}

\onbehalf{on behalf of the GRAMS collaboration}

\affiliation[a]{Department of Physics, Northeastern University, \\
360 Huntington Avenue, Boston, MA 02115, USA}

\emailAdd{zeng.jia@northeastern.edu}
\emailAdd{JCZeng1412@gmail.com}

\usepackage[running]{lineno}

\abstract{The Gamma-Ray and AntiMatter Survey (GRAMS) is a next-generation experiment using a Liquid Argon Time Projection Chamber (LArTPC) detector to measure MeV gamma rays and antiparticles. MeV gamma-ray observations are important for understanding multi-messenger and time-domain astronomy, enabling exploration of the universe's most potent events, such as supernovae and neutron star mergers. Despite the significance of MeV gamma-rays, GRAMS could also explore the so-called 'MeV gap' region to improve MeV gamma-ray measurement sensitivity that was restricted by the challenge of accurately reconstructing Compton events. Aside from gamma-ray detection, the GRAMS proposed method also serves as an antiparticle spectrometer, targeting the low-energy range of cosmic antinuclei measurements. This work will provide updates on the current status and progress towards the prototype balloon flight with a small-scale LArTPC (pGRAMS) scheduled for early 2026, as well as the recent progress on antihelium-3 sensitivity calculation. }

\FullConference{19th International Conference on Topics in Astroparticle and Underground Physics (TAUP2025)\\
24–30 Aug 2025\\
Xichang, China\\}

\begin{document}
\maketitle

\section{Introduction}
In recent astrophysical observations, the Large Area Telescope on the Fermi Gamma-ray Space Telescope (Fermi-LAT) and the Nuclear Spectroscopic Telescope Array (NuSTAR) have deeply explored astrophysical phenomena in the energy domains for hard X-rays (up to 80 keV) and high-energy gamma rays (above 20 MeV), respectively \cite{Atwood_2009, Harrison_2013}. However, gamma rays in the MeV energy range have not yet been well-explored (the so-called “MeV gap”). COMPTEL (The Imaging COMPton TELescope) produced the first catalog of MeV sources, but only approximately 30 objects have been detected \cite{refId0}. MeV gamma-ray astronomy is essential for studying nucleosynthesis, particle acceleration, and violent cosmic events involving black holes and neutron stars. It also plays a key role in multi-messenger astronomy, such as by detecting gamma rays associated with neutron star mergers and gravitational waves.

The Planck experiment provides evidence that 68\% of our universe comprises dark energy, 27\% is dark matter, and 5\% is baryonic matter \cite{refId1}. The existence of dark matter is supported by multiple astronomical observations, including galaxy rotation curves and gravitational lensing in the Bullet Cluster, where two colliding galaxy clusters exhibit clear separation between their mass and baryonic components \cite{10.1046/j.1365-8711.2000.03075.x}. Despite the observational evidence, the nature of dark matter and its interactions with ordinary matter remain poorly understood. Astrophysics experiments to detect dark matter, both directly and indirectly, are ongoing, including satellite missions like the Alpha Magnetic Spectrometer-02 (AMS-02) and the Fermi Gamma-ray Space Telescope (Fermi), as well as balloon-borne experiments, such as the Balloon-borne Experiment with a Superconducting Spectrometer (BESS) and the General Antiparticle Spectrometer (GAPS) \cite{aguilar2002alpha, lubelsmeyer2011upgrade, ajima2000superconducting, hailey2006accelerator}. Gamma-Ray and AntiMatter Survey (GRAMS) represents a novel approach to indirect dark matter detection, specifically optimized for low-energy antinuclei measurements. GRAMS employs liquid argon as its detection medium, enabling the capture of antiparticles and the subsequent decay of their annihilation products. This detection strategy provides a unique window into potential dark matter signatures through antideuteron and antihelium-3 measurements. GRAMS is capable of measuring sub-GeV/n antiprotons, which will provide cross-checks with previous cosmic antiproton measurements as well as instrumental calibration to validate the indirect antinuclei detection method. GRAMS will provide an extensive sensitivity, $\sim$10$^{-6}$ [m$^2$ s sr GeV/n]$^{-1}$, for low-energy antideuteron measurements, as described in the concept paper \cite{aramaki2020dual}. GRAMS is also optimized for the low-energy antihelium-3 measurement with the sensitivity of 1.47 $\times$ 10$^{-7}$ [m$^2$ s sr GeV/n]$^{-1}$ for the 35-day long-duration balloon flights and 1.55 $\times$ 10$^{-9}$ [m$^2$ s sr GeV/n]$^{-1}$ / $3.10\times10^{-10}$ [m$^2$ s sr GeV/n]$^{-1}$, for a 2-year/10-year satellite mission, respectively \cite{ZENG2025103152}.

\section{GRAMS detection concept and science goals}
\begin{figure}[htbp]
\begin{center} 
\includegraphics*[width=15cm]{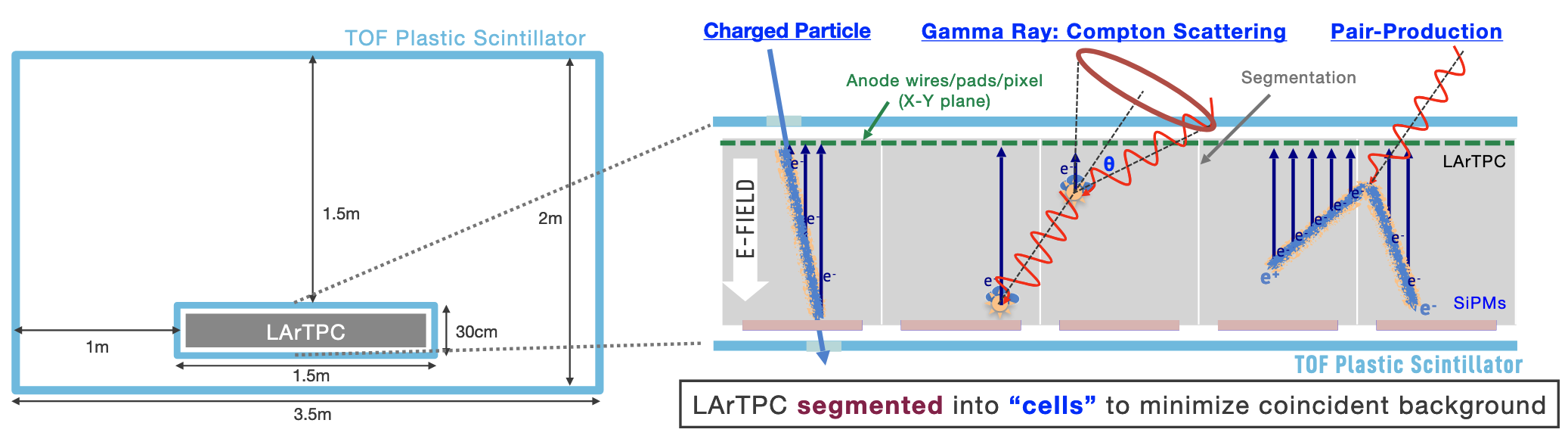}
\end{center}
\caption{GRAMS detection concept. The LArTPC is segmented into “cells” to minimize coincident background events. The right figure shows charged-particle and gamma-ray interactions inside the detector. For the charged-particle case, including antinuclei, the TOF system will measure the velocity and energy deposition. Then, they will slow down as they deposit energy, through ionization, in the LArTPC. The antiparticle will be captured by an argon nucleus, forming an exotic atom. The exotic atom in the excited state will de-excite, emitting Auger electrons and X-rays \cite{aramaki2013measurement}. The antiparticle will eventually be captured by the nucleus and produce annihilation products, including charged pions and protons. The number of pions and protons produced will be related to the number of antinucleons, providing additional information to identify the incoming antiparticle \cite{aramaki2020dual,ZENG2025103152}. For an incoming gamma ray, multi-hit Compton scatterings will be measured and used to reconstruct the Compton ring to determine the source direction. GRAMS has also developed an algorithm to reconstruct the Compton event inside our detector \cite{Takashima:2022, YONEDA2023102765}.}
\end{figure}
The GRAMS detection concept utilizes the combined signals from time-of-flight (TOF) and LArTPC systems to identify and reconstruct events (see Fig.~\ref{schematic}). For the charged particle case, including antinuclei, the TOF system will measure the velocity and energy deposition. Then, they will slow down as they deposit energy, through ionization, in the LArTPC. The antiparticle will be captured by an argon nucleus, forming an exotic atom. The exotic atom in the excited state will de-excite, emitting Auger electrons and X-rays \cite{aramaki2013measurement}. The antiparticle will eventually be captured by the nucleus and produce annihilation products, including charged pions and protons. The number of pions and protons produced here will be related to the number of antinucleons, providing additional information to identify the incoming antiparticle \cite{aramaki2020dual,ZENG2025103152}. Since antiprotons will also form exotic atoms in the LAr detector and generate annihilation products at the stop position, they will contribute the majority of background events for antihelium-3 detection. However, an antihelium-3 nucleus will deposit more energy in the TOF system as it has a charge of -2e. Moreover, since it consists of three antinucleons, the annihilation product profile will be different from an antiproton, where more protons and pions can be generated from the annihilation point \cite{ZENG2025103152}. The atomic x-ray energies are also different between antiproton and antihelium-3 events.

For an incoming gamma ray, multiple Compton scatterings will be measured and used to reconstruct the Compton ring to determine the source direction. GRAMS has also developed algorithms to reconstruct the Compton event  inside our detector \cite{Takashima:2022, YONEDA2023102765}.

\begin{figure}
    \centering
    \includegraphics[width=1\linewidth]{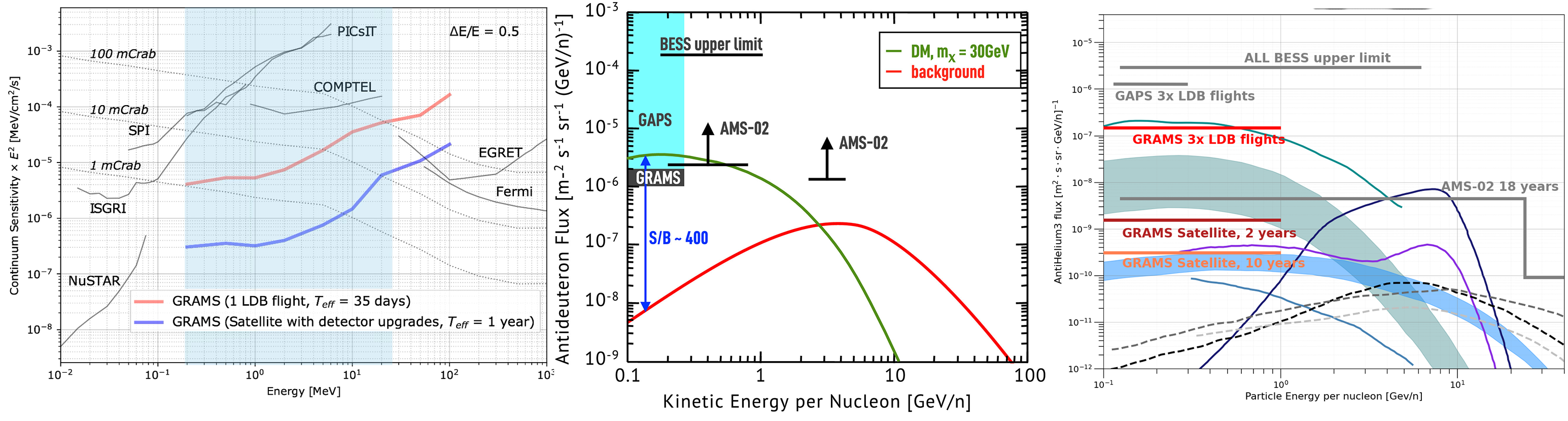}
    \caption{The left figure shows the GRAMS MeV gamma-ray sensitivity projection for a 35 day long-duration balloon flight and a 1-year satellite mission. Both cases would provide more than an order of magnitude improvement over previous missions, including COMPTEL \cite{aramaki2020dual}. The middle plot shows the GRAMS antideuteron sensitivity in comparison with other projects \cite{aramaki2020dual}. The right plot shows the GRAMS antihelium-3 sensitivity in grey lines compared with other experiments. Theoretical primary Dark Matter model predictions are shown in colored lines and areas, while background models are shown in dashed lines\cite{ZENG2025103152}.}
    \label{fig:science}
\end{figure}

GRAMS has proposed a $140\ cm \times 140\ cm \times 20cm$ LArTPC for the science balloon flight. Based on the detection concept described, we have evaluated its capability of detecting MeV gamma rays and antinuclei using GEANT4 Monte Carlo simulations with QGSP\_BERT physics list \cite{AGOSTINELLI2003250, APOSTOLAKIS2009859}. A series of expected scientific sensitivities has been estimated, see Fig.~\ref{fig:science}. The left figure shows the GRAMS MeV gamma-ray sensitivity projection for a 35-day long-duration balloon flight and a 1-year satellite mission. Both cases would provide more than an order of magnitude improvement over previous missions, including COMPTEL \cite{aramaki2020dual}. The middle plot shows the GRAMS antideuteron sensitivity in comparison with other projects \cite{aramaki2020dual}. The right plot shows the GRAMS antihelium-3 sensitivity in grey lines compared with other experiments. Theoretical primary Dark Matter model predictions are shown in colored lines and areas, while background models are shown in dashed lines\cite{ZENG2025103152}.

In conclusion, GRAMS is a pioneering experiment designed for a dual mission: observing MeV gamma rays and searching for dark matter with a LArTPC detector. Its large-scale, cost-effective design will provide over an order of magnitude greater sensitivity to gamma rays and significantly improved sensitivity to antideuterons compared to existing efforts. Currently in development, GRAMS is planned for science flights, following the GAPS and COSI missions.

\section{GRAMS timeline and project status}
\begin{figure}
    \centering
    \includegraphics[width=1\linewidth]{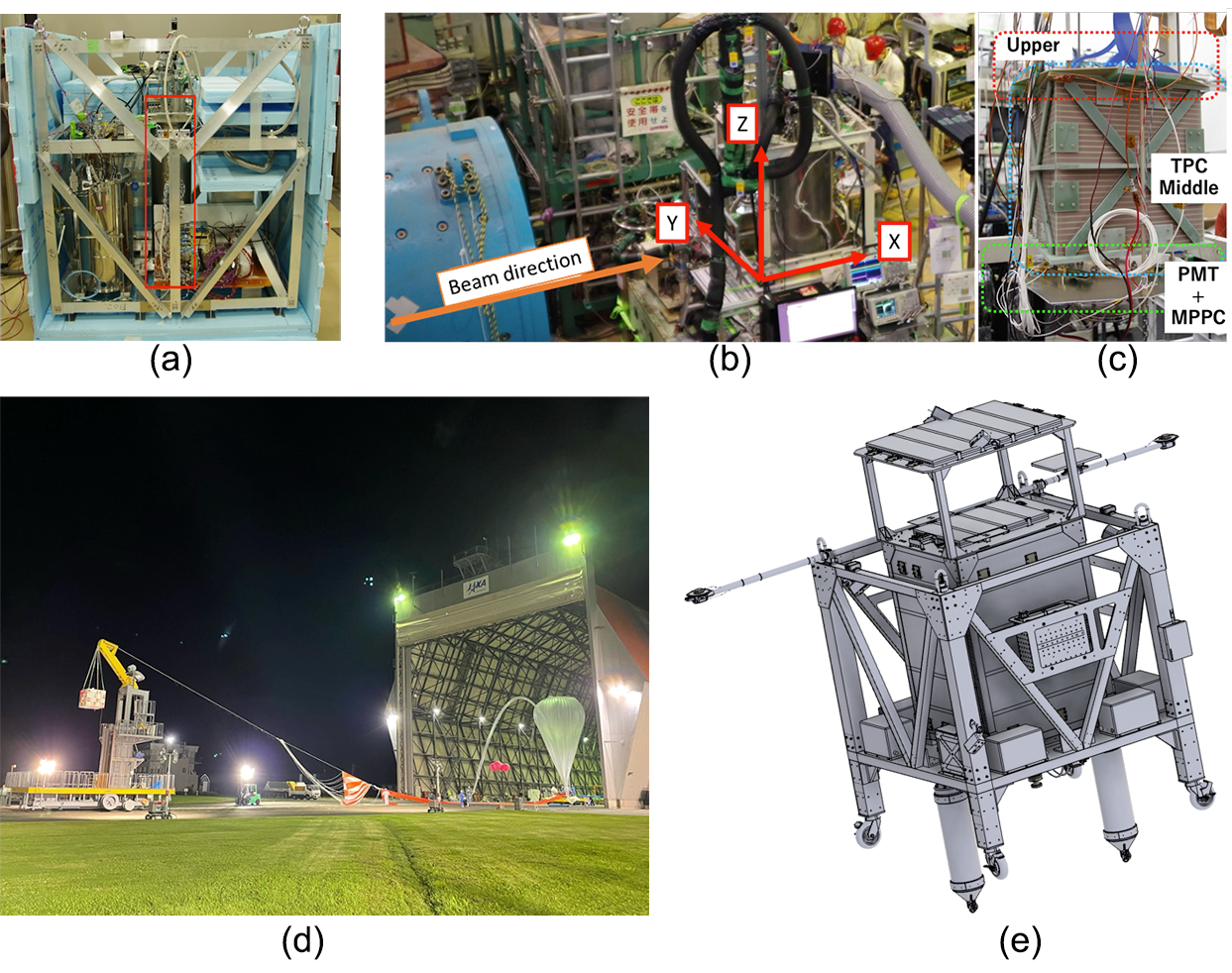}
    \caption{\textbf{(a)} Year 2023 eGRAMS payload. The highlighted red box shows the LArTPC carrier chamber. \textbf{(b)} Year 2025 antiproton beam test at J-PARC. The orange arrow shows the $700MeV/n$ antiproton beam direction. Red xyz arrows show LArTPC chamber orientation. \textbf{(c)} Detector that stays inside the beam test chamber. \textbf{(d)} Year 2023 eGRAMS launch site at Taiki Aerospace Research Field in Hokkaido, Japan. \textbf{(e)} pGRAMS payload design, in preparation for a flight in 2026 at Tucson, Arizona.}
    \label{fig:hardware}
\end{figure}

The GRAMS project was launched in 2017, followed by continuous technological development. GRAMS was funded for an engineer flight in Japan in 2022, and made a successful flight in 2023 \cite{10.1093/ptep/ptae179}. GRAMS then received funding for a prototype balloon flight from the NASA Astrophysics Research and Analysis (APRA) program in 2023. Currently, a smaller-scale detector miniGRAMS with $30\ cm \times 30\ cm \times 20cm$ is being prepared at Northeastern University to be operated for the upcoming pGRAMS mission that is going to launch from Tucson, Arizona. Simultaneously, a similar-sized LArTPC detector was assembled and tested in Japan using a $700MeV/c$ antiproton beam line at the J-PARC T98 experiment\cite{Yano:202503}. Some of the mission-related pictures are shown in Fig.~\ref{fig:hardware}.

\begin{figure}[H]
    \centering
    \includegraphics[width=0.8\linewidth]{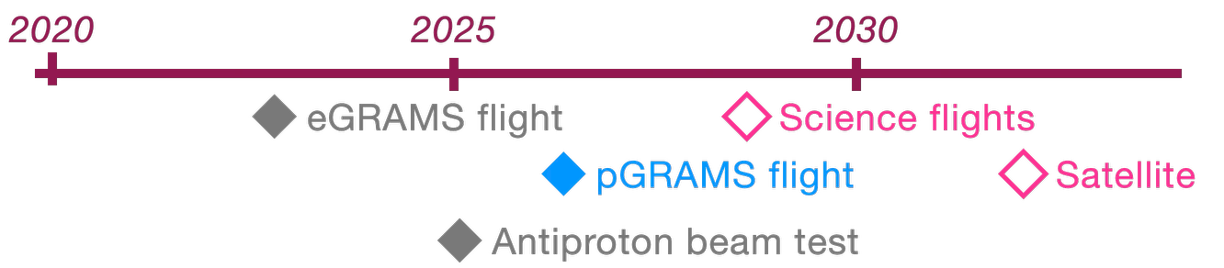}
    \caption{GRAMS timeline}
    \label{fig:timeline}
\end{figure}

In the future, GRAMS will perform science balloon flights and progress to a satellite mission to deliver the science results we proposed and simulated. The planned timeline is shown in Fig.~\ref{fig:timeline}.

\section{Summary}
GRAMS is the first balloon experiment optimized for both MeV gamma-ray observations and an indirect dark matter search with a LArTPC detector. With a cost-effective, large-scale LArTPC detector, GRAMS can provide more than an order of magnitude improvement in sensitivity to MeV gamma rays and significantly improved sensitivity to antinuclei compared to the previous and current ongoing experiments.

\section{Acknowledgments}
This work was supported by the NASA APRA grant, No.22-APRA22-0128 (80NSSC23K1661), and the Alfred P. Sloan Foundation in the US, as well as the JSPS Grant-in-Aid for Scientific Research (A) (22H00133), (B) (22H01252/25K01017) and Challenging Research (Pioneering) (22K18277) in Japan.

\newpage
\section*{GRAMS Collaboration Author List}
\label{sec:authorlist}

\begin{center}
\small
J. Zeng$^{1}$, T. Aramaki$^{1}$, D. Ames$^{1}$, K. Aoyama$^{2,3}$, S. Arai$^{4}$, S. Arai$^{5}$, J. Asaadi$^{6}$, A. Bamba$^{4}$, N. Cannady$^{7}$, P. Coppi$^{8}$, G. De Nolfo$^{7}$, M. Errando$^{9}$, L. Fabris$^{10}$, T. Fujiwara$^{11}$, Y. Fukazawa$^{12}$, P. Ghosh$^{7}$, K. Hagino$^{4}$, T. Hakamata$^{11}$, N. Hiroshima$^{13}$, M. Ichihashi$^{4}$, Y. Ichinohe$^{14}$, Y. Inoue$^{11,15,16}$, K. Ishikawa$^{5}$, K. Ishiwata$^{11}$, T. Iwata$^{4}$, G. Karagiorgi$^{17}$, T. Kato$^{4}$, H. Kawamura$^{11}$, D. Khangulyan$^{18,19}$, J. Krizmanic$^{7}$, J. LeyVa$^{1}$, A. Malige$^{17}$, J. G. Mitchell$^{7}$, J. W. Mitchell$^{7}$, R. Mukherjee$^{20}$, R. Nakajima$^{5}$, K. Nakazawa$^{21}$, H. Odaka$^{11}$, K. Okuma$^{21}$, K. Perez$^{17}$, I. Safa$^{17}$, K. Sakai$^{22}$, M. Sasaki$^{7}$, W. Seligman$^{17}$, J. Sensenig$^{17}$, K. Shirahama$^{11}$, T. Shiraishi$^{23}$, S. Smith$^{24}$, Y. Suda$^{12}$, A. Suraj$^{1}$, H. Takahashi$^{12}$, S. Takashima$^{4}$, T. Tamba$^{4,3}$, M. Tanaka$^{5}$, S. Tandon$^{17}$, R. Tatsumi$^{11}$, J. Tomsick$^{25}$, N. Tsuji$^{23}$, Y. Uchida$^{26}$, Y. Utsumi$^{5}$, S. Watanabe$^{24}$, Y. Yano$^{5}$, K. Yawata$^{27}$, H. Yoneda$^{28, 29}$, K. Yorita$^{5}$, M. Yoshimoto$^{11}$
\end{center}

\vspace{0.5cm}

\noindent

\footnotesize
$^{1}$Department of Physics, Northeastern University, 360 Huntington Avenue, Boston, MA 02115, USA
$^{2}$Department of Physics, Waseda University, 3-4-1 Okubo, Shinjuku-ku, Tokyo 169-8555, Japan
$^{3}$Japan Aerospace Exploration Agency (JAXA), 3-1-1 Yoshinodai, Chuo-ku, Sagamihara City, Kanagawa 252-5210, Japan
$^{4}$Department of Physics, University of Tokyo, Tokyo 113-0033, Japan
$^{5}$Department of Physics, Waseda University, 3-4-1 Okubo, Shinjuku-ku, Tokyo 169-8555, Japan
$^{6}$Department of Physics, University Texas Arlington, 701 South Nedderman Drive, Arlington, TX 76019, USA
$^{7}$NASA Goddard Space Flight Center, 8800 Greenbelt Road, Greenbelt, MD 20771, USA
$^{8}$Department of Astronomy, Yale University, P.O. Box 208101 New Haven, CT 06520-8101, USA
$^{9}$Department of Physics, Washington University at St. Louis, One Brookings Drive, St. Louis, MO 63130-4899, USA
$^{10}$Oak Ridge National Laboratory, 5200, 1 Bethel Valley Rd, Oak Ridge, TN 37830, USA
$^{11}$Department of Earth and Space Science, Graduate School of Science, Osaka University, 1-1 Machikaneyama-cho, Toyonaka, Osaka 560-0043, Japan
$^{12}$Department of Physics, Graduate School of Advanced Science and Engineering, Hiroshima University, 1-3-2, Kagamiyama, Higashi Hiroshima-shi, Hiroshima 739-0046, Japan
$^{13}$Department of Physics, Faculty of Engineering Science, Yokohama National University, Yokohama 240-8501, Japan
$^{14}$RIKEN Nishina Center, Hirosawa 2-1, Wako-shi, Saitama 351-0198, Japan
$^{15}$Interdisciplinary Theoretical \& Mathematical Science Program (iTHEMS), RIKEN, 2-1 Hirosawa, 351-0198, Japan
$^{16}$Kavli Institute for the Physics and Mathematics of the Universe (WPI), UTIAS, The University of Tokyo, 5-1-5 Kashiwanoha, Kashiwa, Chiba 277-8583, Japan
$^{17}$Department of Physics, Columbia University, New York, NY 10027, USA
$^{18}$Key Laboratory of Particle Astrophyics, Institute of High Energy Physics, Chinese Academy of Sciences, 100049 Beijing, China
$^{19}$Tianfu Cosmic Ray Research Center, 610000 Chengdu, Sichuan, China
$^{20}$Department of Physics and Astronomy, Barnard College, 3009 Broadway, New York, NY 10027, USA
$^{21}$Department of Physics, Nagoya University, Furo-cho, Chikusa-ku, Nagoya, Aichi 464-8601, Japan
$^{22}$Enrico Fermi Institute, The University of Chicago, 5640 South Ellis Ave Chicago IL 60637, USA
$^{23}$Faculty of Science, Kanagawa University, 3-27-1, Rokkakubashi, Kanagawa-ku, Yokohama-shi, Kanagawa 221-0802, Japan
$^{24}$Department of Mechanical Engineering, Howard University, 2400 6th St NW, Washington, DC 20059, USA
$^{25}$University of California Berkeley Space Sciences Laboratory, University Avenue and, Oxford St, Berkeley, CA 94720, USA
$^{26}$Department of Physics, Faculty of Science and Technology, Tokyo University of Science, 2641 Yamazaki, Noda, Chiba 278-8510, Japan
$^{27}$Department of Medical Education, National Defense Medical College, 3-2 Namiki, Tokorozawa, Saitama 359-8513, Japan
$^{28}$The Hakubi Center for Advanced Research, Kyoto University, Yoshida Ushinomiyacho, Sakyo-ku, Kyoto 606-8501, Japan
$^{29}$Department of Physics, Kyoto University, Kitashirakawa Oiwake-cho, Sakyo-ku, Kyoto 606-8502,Japan

\vspace{0.5cm}

\noindent
\end{document}